\documentclass{PoS}
\usepackage{wrapfig,rotating}

\title{
Four-fermion production near 
the W-pair production threshold 
}

\ShortTitle{Four-fermion production near the $W$-pair production threshold}

\author{\speaker{Pietro Falgari}
         \\
         Institut f\"ur Theoretische Physik E, RWTH-Aachen\\
         E-mail: \email{falgari@physik.rwth-aachen.de}}

\abstract{I report on recent results for the total
production cross section of the process $e^- e^+ \rightarrow
\mu^- \bar{\nu}_\mu u \bar{d} X$ near the $W$-pair production
threshold up to next-to-leading order in $\Gamma_W/M_W \sim \alpha
\sim v^2$ obtained in the framework of unstable-particle effective
field theory. Remaining theoretical uncertainties and their impact on the
experimental determination of the $W$ mass are discussed.}

\FullConference{8th International Symposium on Radiative Corrections \\
                 October 1-5, 2007\\
                 Florence, Italy}

\newcommand{\braket}[1]{\langle #1 \rangle}
\newcommand{\Tprod}[1]{{\mathrm T}\lbrack #1 \rbrack}

\begin{document}

\section{Introduction}

An accurate measurement of the $W$ mass is of primary interest
for precision tests of the Standard Model and for search of 
New-Physics effects through virtual-particle exchange. The total error
on $M_W$ could be lowered to $6 \,\, \mbox{MeV}$ by measuring the
four-fermion production cross section near the $W$-pair production
threshold \cite{Wilson:2001aw} at a future International Linear
Collider (ILC), provided that the theoretical uncertainties are well
below $1 \%$. This is a difficult task, requiring
gauge-invariant inclusion of finite-width effects and
calculation of QCD and electroweak radiative corrections to the
full $2 \rightarrow 4$ process. Previous NLO calculations 
in the double-pole approximation
\cite{Denner:2000bj} were supposed to break down near threshold
for kinematical reasons. The recent computation of the complete
NLO corrections to $e^- e^+ \rightarrow 4 f $ in the
complex-mass scheme \cite{Denner:2005fg} is valid both near threshold
and in the continuum, but is technically difficult, requiring the
computation of one-loop six-point functions.

Here I present NLO results for the total cross section of the process
\begin{equation}\label{eq:4fermions}
e^- e^+ \rightarrow \mu^- \bar{\nu}_\mu u \bar{d} X
\end{equation}
near the $W$-pair production 
threshold \cite{Beneke:2007zg}
computed 
with effective field theory (EFT) techniques
\cite{Chapovsky:2001zt,Beneke:2004xd,Beneke:2004km}.
Section \ref{sec:uptheory} reviews briefly the formalism, while the 
calculation of the Born cross section and of radiative
corrections is outlined in Sections \ref{sec:tree} and
\ref{sec:radiative}. 
Section \ref{sec:results}
presents numerical results together with an estimate of the
remaining theoretical uncertainties and a comparison with
\cite{Denner:2005fg}.

\section{Unstable-particle effective field theory}
\label{sec:uptheory}

The EFT approach
\cite{Beneke:2004km} exploits the
hierarchy of scales $M \Gamma \ll M^2$ 
which characterizes processes involving
unstable particles, 
$M$ and $\Gamma$ being the mass and width of the
intermediate resonance. The 
degrees of freedom of the
full theory are
classified according to their scaling into 
short-distance ($k^2 \sim M^2$) and long-distance
($k^2 \lesssim M \Gamma$) modes. The fluctuations at the small
scale (resonant particles, soft and Coulomb photons,...) 
represent the field content of the
effective Lagrangian $\cal{L}_{\mbox{\scriptsize eff}}$. ``Hard''
fluctuations with $k^2 \sim M^2$ are not part of the effective
theory and are integrated out. Their effect is 
included in $\cal{L}_{\mbox{\scriptsize eff}}$ through short-distance
matching coefficients, computed in standard \emph{fixed-order}
perturbation theory. The systematic inclusion of finite-width
effects is relevant
for modes with virtuality $k^2 \lesssim M
\Gamma$ and is obtained through complex short-distance  
coefficients in $\cal{L}_{\mbox{\scriptsize eff}}$ \cite{Beneke:2004km}.

The specific process (\ref{eq:4fermions}) is primarily  
mediated by production of a pair of resonant $W$s.
The total cross section is extracted from appropriate cuts of
the forward-scattering amplitude \cite{Beneke:2007zg}, which after
integrating out the hard modes with $k^2 \sim M_W^2$ reads 
\cite{Beneke:2004km}
\begin{eqnarray}
\label{eq:master}
&& i {\cal A} =\sum_{k,l} \int d^4 x \,
\braket{e^- e^+ |
\Tprod{i {\cal O}_p^{(k)\dagger}(0)\,i{\cal O}_p^{(l)}(x)}|e^- e^+}
+ \sum_{k} \,\braket{e^- e^+|i {\cal O}_{4e}^{(k)}(0)|e^- e^+}.
\\[-0.5cm]
\nonumber
\end{eqnarray}
The operators ${\cal O}_p^{(l)}$ (${\cal O}_p^{(k)\dagger}$)
in the first term on the right-hand side of (\ref{eq:master}) 
produce (destroy) a pair
of non-relativistic resonant $W$ bosons. The second term accounts
for the remaining non-resonant contributions. The computation of
$\cal{A}$ is split into the determination of the matching
coefficients of the operators ${\cal O}_p^{(l)}$, ${\cal
O}_{4e}^{(k)}$ and the calculation of the matrix elements in
(\ref{eq:master}). Both quantities are computed as power series
in the couplings $\alpha, \, \alpha_s$, the ratio $\Gamma_W/M_W$ and the 
non-relativistic velocity of the intermediate resonant $W$ pair 
$v^2 \equiv (\sqrt{s}-2 M_W)/(2M_W)$, collectively referred 
to as $\delta \sim \alpha_s^2 \sim \alpha \sim \Gamma_W/M_W \sim v^2$.

The effective Lagrangian describing the non-relativistic $W$
bosons up to NLO in $\delta$ is
\cite{Beneke:2004xd}
\begin{equation}
{\cal L}_{\rm NRQED} = \sum_{a=\mp} \left[\Omega_a^{\dagger i} \left(
i D^0 + \frac{\vec{D}^2}{2 {M}_W} - \frac{\Delta}{2} \right)
\Omega_a^i
+  \Omega_a^{\dagger i}\,
\frac{(\vec{D}^2-M_W \Delta)^2}{8 M_W^3}\,
\Omega_a^i\right].
\label{LNR}
\end{equation}
$\Delta$ is the matching coefficient $\Delta \equiv
(\bar{s}-M_W^2)/M_W$, where $\bar{s}$ is the complex pole of the
$W$ propagator. 
The field $\Omega_\pm^i=\sqrt{2 M_W} W_{\pm}^i$ describes the 
three physical polarizations of
non-relativistic $W$s, and the covariant derivative
$D_\mu \Omega_\pm=(\partial_\mu \mp i e A_\mu) \Omega_\pm$ contains
the interaction of the resonant fields $\Omega_\pm$ with 
\emph{soft} and \emph{potential} photons (see Section 
\ref{sec:radiative}).
To complete $\mathcal{L}_{\mbox{\scriptsize eff}}$ one has to 
add to (\ref{LNR}) the effective production vertices
${\cal O}_p^{(l)}$ and the four-fermion operators ${\cal
O}_{4e}^{(k)}$ with the corresponding matching coefficients computed
to the desired order in $\delta$. These are presented in Sections
\ref{sec:tree} and \ref{sec:radiative}.

\section{EFT approximation to the Born cross section}\label{sec:tree}

The lowest-order production operator of two non-relativistic resonant $W$s 
is \cite{Beneke:2004xd}
\begin{equation}
{\cal O}_p^{(0)} = \frac{\pi\alpha_{ew}}{M_W^2}
\left(\bar{e}_{c_2,L} (\gamma^i n^j+\gamma^j n^i) e_{c_1,L}
\right) \left(\Omega_-^{\dagger i} \Omega_+^{\dagger j}\right) .
\label{LPlead}
\end{equation}
Its matching coefficient is extracted from the \emph{on-shell}
process $e^- e^+ \rightarrow W^- W^+$, where ``on-shell'' means
$k^2=\bar{s}$.
The four-fermion operators ${\cal O}_{4e}^{(k)}$ do not contribute 
to $\mathcal{A}$ at this order, and the forward-scattering amplitude is simply
\begin{equation} \label{eq:LOforward}
i \mathcal{A}^{(0)} = \int d^4 x \langle e^- e^+|T[i
\mathcal{O}^{(0)\dagger}_p(0) i \mathcal{O}_p^{(0)}(x)]|e^-
e^+\rangle =
\parbox[h]{2.5 cm}{\includegraphics[width=\linewidth]{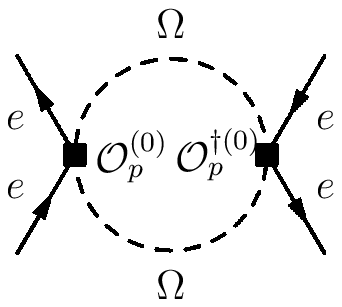}}
=-\frac{i \pi\alpha^2}{s_w^4} \sqrt{-\frac{E+i
\Gamma_W^{(0)}}{M_W}},
\end{equation}
with $E=\sqrt{s}-2 M_W$ and $s_w=\sin \theta_W$.
The total cross section for (\ref{eq:4fermions}) is extracted from
appropriate cuts of (\ref{eq:LOforward}). At lowest order this is
correctly done by multiplying the imaginary part of
$\mathcal{A}^{(0)}$ with the LO branching ratios of the
decays $W^-\rightarrow \mu^-\bar{\nu}_\mu$, $W^+ \rightarrow u
\bar{d}$, so that $\sigma^{(0)} = \frac{1}{27 s}\,\mbox{Im}\,{\cal A}^{(0)}$.
 
Beyond the leading term $\sigma^{(0)}$
there are contributions which can be identified with terms of the
expansion in $\delta$ of a full-theory Born result 
computed with a fixed-width prescription. The first class of corrections arises
from \emph{four-electron operators} in
(\ref{eq:master}). The imaginary part of their matching coefficients
are extracted from suitable cuts of \emph{hard} two-loop SM diagrams
\cite{Beneke:2007zg}:
\begin{equation}\label{eq:4electrons}
\parbox[h]{9 cm}{\includegraphics[width=\linewidth]{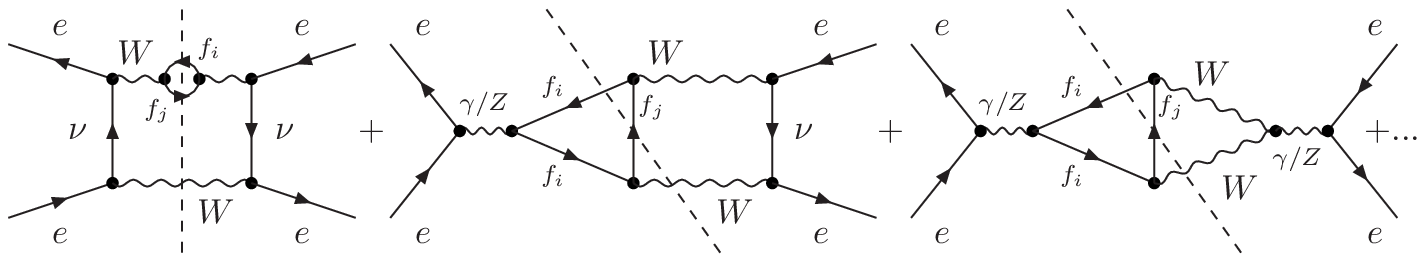}}\Rightarrow
\parbox[h]{1.9 cm}{\includegraphics[width=\linewidth]{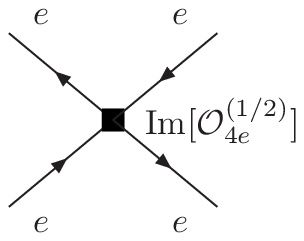}}
\end{equation}
Compared to the LO cross section $\sigma^{(0)} \sim
\alpha^2 \sqrt{\delta}$ the new term is suppressed by
$\alpha/\sqrt{\delta} \sim \sqrt{\delta}$ and is denoted as
``$\sqrt{\mbox{N}}\mbox{LO}$''. True NLO contributions to
$\mathcal{A}^{(0)}$ arise from \emph{higher-dimension 
production operators} and \emph{propagator corrections}. The
former come from the matching of the effective theory on the
on-shell process $e^- e^+ \rightarrow W^- W^+$ at order $v$
($\mathcal{O}^{(1/2)}_p$) and $v^2$ ($\mathcal{O}^{(1)}_p$)
\cite{Beneke:2004xd}. The latter correspond to the term
$(\vec{\partial}^2-M_W \Delta)^2/(8 M_W^3)$ in (\ref{LNR}). 
A comparison of the EFT Born approximations with the full 
result computed with 
Whizard \cite{Kilian:2001qz} shows a good convergence of the 
series \cite{Beneke:2007zg}. However partial inclusion of
N$^{3/2}$LO corrections is necessary to obtain an agreement of
$\sim 0.1 \%$ at $170 \,\, \mbox{GeV}$ and $\sim 10\%$ at $155
\,\, \mbox{GeV}$ \cite{Beneke:2007zg}. 

\section{Radiative corrections}\label{sec:radiative}

A complete NLO prediction must include radiative corrections to 
the Born result. These are electroweak and QCD corrections to 
the matching coefficient of ${\cal O}_p^{(0)}$ and loop contributions 
to the EFT matrix elements. At NLO the flavor-specific final state 
is selected by multiplying the total cross section with NLO
branching ratios. The $O(\alpha)$ correction to the matching
coefficient of (\ref{LPlead}) is obtained from the one-loop
amplitude of $e^- e^+\rightarrow W^- W^+$.  
Many of the 180 one-loop diagrams do not contribute 
due to threshold kinematics and the result reads \cite{Beneke:2007zg}:
\begin{equation}
C^{(1)}_p=\frac{\alpha}{2 \pi} \left[\left(-\frac{1}{\varepsilon^2}
-\frac{3}{2 \varepsilon}\right) \left(-\frac{4 M_W^2}{\mu^2}
\right)^{-\varepsilon}+c_p^{(1,\mbox{\tiny fin})}\right]
\end{equation}

The one-loop corrections to the matrix elements arise from
exchange of \emph{potential} ($(q_0,|\vec{q}|) \sim M_W (\delta, 
\sqrt{\delta}) $) and \emph{soft} ($(q_0,|\vec{q}|) \sim M_W (\delta, 
\delta) $) photons. Loops containing $n$ potential photons are
enhanced by inverse powers of $v$, 
$\Delta \mathcal{A} \sim \mathcal{A}^{(0)} \alpha^n v^{-n} 
\sim \mathcal{A}^{(0)} \alpha^{n/2}$, so that the first and second
Coulomb corrections must be included in a NLO calculation. Near
threshold they amount respectively to $\sim 5 \%$ and 
$\sim 0.2 \%$  of $\sigma^{(0)}$ \cite{Beneke:2007zg}.

Two-loop diagrams with soft photons connecting different hard 
subprocesses of (\ref{LPlead}) give the so-called 
\emph{non-factorizable} corrections. 
As a consequence of the residual gauge-invariance of 
$\mathcal{L}_{\mbox{\scriptsize eff}}$, and in agreement with previous results 
\cite{Fadin:1993dz}, only the initial-initial state interferences survive:
\begin{equation}
\parbox[h]{4.3 cm}{\includegraphics[width=\linewidth]{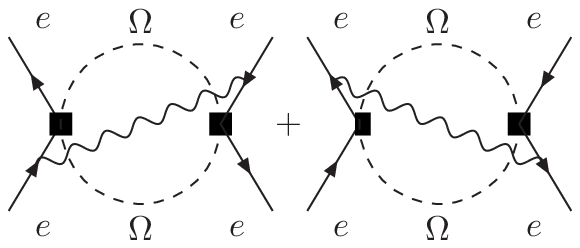}}=
\frac{4 \pi^2 \alpha^2}{s_w^4 M_W^2}\frac{\alpha}{\pi} \int
\frac{d^d r}{(2 \pi)^d} \frac{1}{\eta_{-} \eta_{+}}
\left[\left(\frac{1}{\varepsilon^2}+\frac{5}{12} \pi^2 \right)
\left(-\frac{2 \eta_-}{\mu}\right)^{-2 \varepsilon}\right],
\end{equation}
with $\eta_-=r_0-\frac{|\vec{r}|^2}{2 M_W}+i \frac{\Gamma_W^{(0)}}{2}$ and
$\eta_+=E-r_0-\frac{|\vec{r}|^2}{2 M_W}+i \frac{\Gamma_W^{(0)}}{2}$.

\section{Results and remaining theoretical uncertainties}\label{sec:results}

\begin{wrapfigure}{r}{6 cm}
\includegraphics[width=\linewidth]{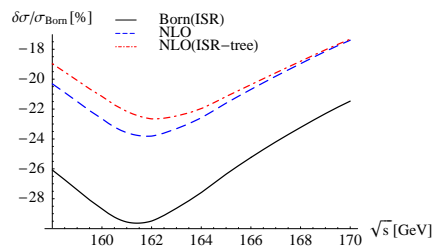}
\caption{Size of the relative NLO corrections for different
implementations of ISR}
\label{fig:relativecorrections}
\end{wrapfigure}

Because of the approximation $m_e=0$, the sum of the corrections 
calculated in Section \ref{sec:radiative} is not
infrared safe, containing uncanceled $\varepsilon$-poles. 
The result should be convoluted with $\overline{\mbox{MS}}$
electron distribution functions after minimal subtraction of the
pole. Since the distributions available in the literature are
computed in a different scheme, which assumes $m_e$ as infrared
regulator, it is more convenient to convert our result from 
$\overline{\mbox{MS}}$ to this scheme. This is done by adding 
contributions from the \emph{hard-collinear} ($k^2 \sim
m_e^2$) and \emph{soft-collinear} ($k^2 \sim m_e^2
\frac{\Gamma_W}{M_W}$) regions. These cancel the 
$\varepsilon$-poles, but introduce large logs of $2 M_W/m_e$
\cite{Beneke:2007zg}.
The large logs are resummed by convoluting the NLO cross section 
with the structure functions $\Gamma^{\mbox{\tiny LL}}_{ee}$
used in \cite{Denner:2000bj} after
subtracting the double counting terms \cite{Beneke:2007zg}. 
Since only leading logs are resummed in $\Gamma^{\mbox{\tiny LL}}_{ee}$,
one can equivalently choose to convolute only the Born cross
section with the structure functions, as done for example
in \cite{Denner:2005fg}, the difference being formally NLL. 
Fig. \ref{fig:relativecorrections} shows the percentual correction
to the Born result due to initial-state radiation 
alone (solid black), full NLO corrections with ISR improvement of the Born 
cross-section only (dot-dashed red), and complete NLO corrections with 
full ISR improvement (dashed blue). The contribution of genuine
electroweak and QCD corrections amounts to $\sim 8 \%$ at
threshold. It must also be noted that the difference between the two
implementations of ISR is numerically important, reaching $\sim 2 \%$
at threshold. A comparison of the
EFT approximation with \cite{Denner:2005fg} reveals a discrepancy
which is never larger than $\sim 0.6 \%$ in the range $161\,
\mbox{GeV}<\sqrt{s} <170 \, \mbox{GeV}$. More precisely we have
for the full calculation $\sigma_{\mbox{\scriptsize 4f}}(161
\, \mbox{GeV})= 118.12(8)\,\mbox{fb}$, $\sigma_{\mbox{\scriptsize 4f}}(170
\, \mbox{GeV})=401.8(2)\,\mbox{fb}$ \cite{Denner:2005fg}, while 
in the EFT one obtains $\sigma_{\mbox{\scriptsize EFT}}(161
\, \mbox{GeV})= 117.38(4)\,\mbox{fb}$, $\sigma_{\mbox{\scriptsize EFT}}(170
\, \mbox{GeV})=399.9(2)\,\mbox{fb}$ \cite{Beneke:2007zg}.

The dominant remaining theoretical uncertainty comes from an
incomplete NLL treatment of ISR. This translates
into an uncertainty on the $W$ mass of $\sim 31 \, \mbox{MeV}$
\cite{Beneke:2007zg}. Further uncertainties come from N$^{3/2}$LO 
corrections in the EFT. The missing $O(\alpha)$ corrections 
to the four-electron operator (\ref{eq:4electrons}), which
are included in \cite{Denner:2005fg}, contributes 
an estimated uncertainty of $\sim 8 \, \mbox{MeV}$ \cite{Beneke:2007zg}, 
while interference of potential and soft photon exchange accounts for 
additional $\sim 5 \, \mbox{MeV}$ \cite{Beneke:2007zg}. 
This means that with a NLL treatment of initial-state radiation, 
which seems realistically achievable in the near future,
and further inputs from \cite{Denner:2005fg} the total theoretical
error on $M_W$ could be reduced to the level required for
phenomenological applications at linear colliders.

\section*{Acknowledgments}

I thank M. Beneke, C. Schwinn, A. Signer and G. Zanderighi for the
collaboration on \cite{Beneke:2007zg} and for comments on the manuscript.


\begin{thebibliography}{99}

  \bibitem{Wilson:2001aw}
  G.~Wilson,
  \newblock in {\em 2nd ECFA/DESY Study}, pp. 1498--1505,
  \newblock Desy LC note LC-PHSM-2001-009.

  \bibitem{Denner:2000bj}
  A.~Denner, S.~Dittmaier, M.~Roth and D.~Wackeroth,
  \newblock {\em Nucl. Phys.} {\bf B587}, 67 (2000), [hep-ph/0006307];
  \newblock {\em Phys. Lett.} {\bf B475}, 127 (2000), [hep-ph/9912261];
  W.~Beenakker, F.~A. Berends and A.~P. Chapovsky,
  \newblock {\em Nucl. Phys.} {\bf B548}, 3 (1999), [hep-ph/9811481].

  \bibitem{Denner:2005fg}
  A.~Denner, S.~Dittmaier, M.~Roth and L.~H. Wieders,
  \newblock {\em Nucl. Phys.} {\bf B724}, 247 (2005), [hep-ph/0505042];
  \newblock {\em Phys. Lett.} {\bf B612}, 223 (2005), [hep-ph/0502063].

  \bibitem{Beneke:2007zg}
  M.~Beneke, P.~Falgari, C.~Schwinn, A.~Signer, G.~Zanderighi,
  \newblock {\em Nucl. Phys.} {\bf B792}, 89 (2008) 
  [arXiv:0707.0773 [hep-ph]]; 
  C.~Schwinn, \newblock [arXiv:0708.0730 [hep-ph]].

  \bibitem{Chapovsky:2001zt}
  A.~P. Chapovsky, V.~A. Khoze, A.~Signer and W.~J. Stirling,
  \newblock {\em Nucl. Phys.} {\bf B621}, 257 (2002), [hep-ph/0108190].

  \bibitem{Beneke:2004xd}
  M.~Beneke, N.~Kauer, A.~Signer and G.~Zanderighi,
  \newblock {\em Nucl. Phys. Proc. Suppl.} 
  {\bf 152}, 162 (2006), [hep-ph/0411008].
  
  \bibitem{Beneke:2004km}
  M.~Beneke, A.~P. Chapovsky, A.~Signer and G.~Zanderighi,
  \newblock {\em Nucl. Phys.} {\bf B686}, 205 (2004), [hep-ph/0401002].

  \bibitem{Kilian:2001qz}
  W.~Kilian,
  \newblock in {\em 2nd ECFA/DESY Study}, pp. 1924--1980,
  \newblock DESY LC-Note LC-TOOL-2001-039.

  \bibitem{Fadin:1993dz}
  V.~S. Fadin, V.~A. Khoze and A.~D. Martin,
  \newblock {\em Phys. Rev.} {\bf D49}, 2247 (1994);
  K.~Melnikov and O.~I. Yakovlev,
  \newblock {\em Phys. Lett.} {\bf B324}, 217 (1994), [hep-ph/9302311].

\end{thebibliography}
\end{document}